\documentclass[journal,draftcls,onecolumn,12pt,twoside]{IEEEtranTCOM}
\usepackage{threeparttable}
\usepackage{verbatim}
\usepackage{cite}
\usepackage{mathtools}
\makeatother
\usepackage{amsmath,amssymb,placeins}
\usepackage{algorithmic}
\usepackage{graphicx}
\usepackage{textcomp}
\usepackage[mathscr]{euscript}
\usepackage{lipsum}
\usepackage{relsize}
\usepackage{comment}
\usepackage{cuted}
\usepackage{bm}
\usepackage{multicol}
\usepackage{array}
\usepackage{makecell}
\usepackage{diagbox}
\usepackage[para]{footmisc}
\usepackage{tabularx}

\usepackage[mathscr]{euscript}
\usepackage[usenames]{color}
\usepackage{tablefootnote}
\usepackage{amsfonts}
\usepackage{latexsym}
\usepackage{subfigure}
\usepackage{caption}
\usepackage{float}
\usepackage{multirow}
\usepackage{epstopdf}

\newtheorem{theorem}{{{\textit{Theorem}}}}

\newtheorem{lemma}{{{\textit{Lemma}}}}
\newtheorem{corollary}{{{{\textit{Corollary}}}}}

\newtheorem{definition}{{{\textit{Definition}}}}

\newtheorem{remark}{{{\textit{Remark}}}}

\newtheorem{example}{{{\textit{Example}}}}

\newtheorem{case}{{{\textit{Case}}}}
\normalsize

\ifCLASSINFOpdf
\else
\fi

\begin{document}
\title{A Direct Construction of Complete Complementary
Code with Zero Correlation Zone property for
Prime-Power Length}
\author{Nishant Kumar, Sudhan~Majhi, \IEEEmembership{Senior Member, IEEE},~and~A.K. Upadhyay}
\IEEEpeerreviewmaketitle
\maketitle
\begin{abstract} 
In this paper, we propose a direct construction of a novel type of code set, which has combined properties of complete complementary code (CCC) and zero-correlation zone (ZCZ) sequences and called it complete complementary-ZCZ (CC-ZCZ) code set. The code set is constructed by using multivariable functions. The proposed construction also provides Golay-ZCZ codes with new lengths, i.e., prime-power lengths. 
 The proposed Golay-ZCZ codes are optimal and asymptotically optimal for binary and non-binary cases, respectively, by \emph{Tang-Fan-Matsufuzi} bound. Furthermore, the proposed direct construction provides novel ZCZ sequences of length $p^k$, where $k$ is an integer $\geq 2$. 
We establish a relationship between the proposed CC-ZCZ code set and the first-order generalized Reed-Muller (GRM) code, and proved that both have the same Hamming distance. We also counted the number of CC-ZCZ code set in first-order GRM codes. 
The column sequence peak-to-mean envelope power ratio (PMEPR) of the proposed CC-ZCZ construction is derived and compared with existing works. The proposed construction is also deduced to Golay-ZCZ  and ZCZ sequences which are compared to the existing work. The proposed construction generalizes many of the existing work. 

\end{abstract}
\section{Introduction}\label{sec:intro}
\IEEEPARstart{G}{OLAY} was the first to introduce Golay complementary pair (GCP) in 1949 but defined formally in his article “Complementary Series” in 1960 \cite{Golay}. GCPs show their applications in many areas such as infrared multislit spectrometry, orthogonal frequency-division multiplexing (OFDM) and many others. The idea of GCP was further extended to Golay complementary sets (GCSs) \cite{TsLi}. A relationship between GCP of length in the form $2^m$ and generalized Boolean function (GBF) has been established in \cite{DaJe}. Later in \cite{Paterson}, Paterson \emph{et al.} obtained GCSs of polyphase sequences from cosets of the first-order generalized Reed-Muller (GRM) code. Further, an upper bound on peak-to-mean envelope power ratio (PMEPR) for second-order cosets of this code was obtained. In \cite{HaSu}, the idea of GCSs was further extended to complete complementary codes (CCCs) which is a collection of GCSs with some extra aperiodic cross-correlation
property. Additionally, CCC has found application in multi-carrier code division multiple access (MC-CDMA) systems to eliminate multiple-access interference (MAI) for multiple users over asynchronous environment \cite{ShBe,HsJu}. CCCs are also utilized in optimal channel estimation for multiple-input multiple-output (MIMO) frequency-selective fading channels \cite{wang}, MIMO radar \cite{shufeng, tang}, cell search in OFDM systems \cite{min}, and data hiding\cite{kojima}. A detailed study on GCSs and CCCs can be carried out in \cite{DaJe,Paterson,ArAk,SaLiMa,wang2019new,wang2017A_method,wang2021new,chen2008complete,kumar2021direct}.
\par 
In OFDM systems, GCSs are widely used to control PMEPR \cite{DaJe}, an upper bound on PMEPR was obtained in \cite{Paterson}. By using the good aperiodic correlation properties of GCP and GCSs, these can be used for asynchronous CDMA  system, channel estimation \cite{SpPr}, synchronization \cite{GrMa}, signal detection etc. 
Quasi-synchronous CDMA (QS-CDMA) system is proposed to reduce the impact of synchronization accuracy in communication systems \cite{WaHu}. QS-CDMA system can eliminate co-channel interference for multiuser environment when synchronization error is controlled within a specific range \cite{WaHu}.

Zero correlation zone (ZCZ) sequences have ideal periodic correlation property inside a zone about the origin. Therefore, they can reduce MAI and multipath interference (MPI) \cite{LoPi}, and hence they are used to take benefit of quasi synchronization in the QS-CDMA system. 
They are also used in radar \cite{HeRe}, channel estimation \cite{HuLiu,YaWu,YuWa}, pilot design \cite{ZhCh,ZhZe}, multicode MIMO system \cite{Yang}, and as a training sequence in  MIMO-OFDM systems\cite{ZhCh}. The ZCZ sequences can achieve optimal channel estimation performance under the condition that all the received signals are quasi-synchronous within the ZCZ \cite{HeHu,FrC}.

The relation between GCSs and ZCZ sequences was first established in \cite{DeFa}. Later, numerous ZCZ sequences are constructed by using mutually orthogonal GCSs (MOGCS) \cite{RaCh,TaFa,LiCh}. While most of the ZCZ sequence constructions need predefined sequences like perfect sequences and use unitary matrices\cite{LiCh}, interleaving technique \cite{ZhTa} and Hadamard matrices \cite{ToNaSu}. The direct constructions of ZCZ sequences based on GBFs are provided in \cite{TaYu,LiGuPa,NiSu}. The connection between GRM codes and ZCZ sequences is also provided in \cite{LiGuPa,NiSu}. The GCSs and ZCZ sets are characterized by their aperiodic auto-correlation and periodic correlation properties, respectively. The authors in \cite{GoHu,GonHu}, investigated the periodic auto-correlation behaviour of individual Golay sequences in 2013. Specifically, they presented two constructions of Golay sequences of length $2^m$ each displaying a periodic zero auto-correlation zone (ZACZ) of $2^{m-2}$, and $2^{m-3}$, respectively, around the
in-phase position. 
Motivated by the work in \cite{GoHu,GonHu},
the authors in \cite{ChWu} studied the periodic zero cross-correlation zone (ZCCZ) of the GCSs presented in \cite{chen2008complete} and provided a relation between ZCZ sets and GCSs and proposed a construction of a new class of sequence sets and named it Golay-ZCZ sequence sets. Golay-ZCZ sequence set consists of sequences having their aperiodic auto-correlation sum zero except at the zero shift and ideal periodic cross-correlations and auto-correlations within the ZCZ. Precisely, a Golay-ZCZ sequence set is a ZCZ sequence set which is also a GCS. Therefore, from the application point of view, Golay-ZCZ sequence sets are used to obtain both the advantages in a single code. For an instant, ZCZ sequences are used as pilots or training sequences in OFDM systems \cite{ZhZe,ZhCh} but it is not utilized for PMEPR reduction. Since Golay-ZCZ sequence sets are themselves GCSs, they can be utilized as training sequences as well as for PMEPR reduction. Therefore, the proposed code provides low PMEPR when employed in OFDM systems due to the good auto-correlation properties of GCSs. Besides, the Golay-ZCZ sets possess zero auto-correlations and cross-correlations within their ZCZ and hence have the potential
application to synchronization. 
Inspired by the Golay-ZCZ work given in \cite{ChWu}, the authors in \cite{WaHu} given construction of Golay-ZCZ sequence sets which have large ZCZ width. 
The length of proposed Golay-ZCZ sequence sets in \cite{WaHu} and \cite{ChWu} is limited to power-of-two. Continuing in this chain, recently in 2021, the authors in \cite{Gu2021asymptotically} provided a construction of Golay-ZCZ sequences of length $M^2N$ using an CCC of length $N$ and flock size $M$. Furthermore, in \cite{GuZh}, they also constructed Golay-ZCZ complementary pair of length $4N$, where $N$ is a positive integer and provided an open problem to construct Golay-ZCZ sequence set of non-power-of-two length as \emph{\textquotedblleft An interesting future work will be to design Golay-ZCZ sequence sets, consisting sequences of length non-power-of-two, which have periodic ZACZ\footnote{zero auto-correlation zone} and ZCCZ\footnote{zero cross-correlation zone} around the in-phase position"}. Furthermore, they also extended this work to two-dimension \cite{Gu2021New}.
\par
Motivated by the above works, and open problem provided in \cite{GuZh}, in this paper, we provide a construction of novel type of code set called CC-ZCZ code set of prime-power length in which each code is Golay-ZCZ sequence set  having non-power-of-two length sequences which settles the open problem provided by in \cite{GuZh}. The proposed $p$-ary Golay-ZCZ sequence sets are asymptotically optimal with respect to \emph{Tang-Fan-Matsufuji} bound and optimal for binary case. The proposed CC-ZCZ code sets are also seen as second-order cosets of first-order GRM codes. Since first-order GRM codes enjoy high Hamming distance, we prove that the proposed Golay-ZCZ sequence set in CC-ZCZ code set has the same Hamming distance as GRM codes. Additionally, we also count the number of cosets of GRM codes, which is further used to count the total number of CC-ZCZ code sets and Golay-ZCZ sequence sets. 
The column sequence PMEPR of the proposed CC-ZCZ construction is reduced. The row and column sequence PMEPR for Golay-ZCZ and CCC are also been reduced over  the existing work \cite{ChWu} and \cite{SaLiMa}, respectively. The proposed construction can generate new ZCZ sequences of length $p^k$, where $k$ is a positive integer $\geq 2$. The proposed construction also contributes Golay-ZCZ sequence sets and ZCZ sequences which are again compared to the existing state-of-the-art. 
\par
The remainder of the paper is structured as follows. We provide some basic notations and definitions in Section II. In Section III, we provide the construction of CC-ZCZ code sets and provide examples in support of our construction. In Section IV, a relation between proposed CC-ZCZ code set and GRM codes is established. Moreover, the number of CC-ZCZ code set in a second-order coset of GRM code and Hamming distance are provided. In Section V, we discuss column sequence PMEPR of constructed CC-ZCZ code based MC-CDMA system. Section VI provides another result that increases the number of CC-ZCZ code sets. In section VII, we defend the work's novelty by comparing it to existing works in the literature. Lastly, we provide conclusion in Section VIII.

\section{Preliminary}
\subsection{Definition and Correlation Functions}
Let $\mathbf{a}=(a_0,a_1,\hdots, a_{L-1})$ and $\mathbf{b}=(b_0,b_1,\hdots, b_{L-1})$ be two $L$-length complex-valued sequences. For an integer $\tau$, define
\begin{equation}\label{equ:cross}
\mathcal{A}(\mathbf{a}, \mathbf{b})(\tau)=
\begin{cases}
\sum_{i=0}^{L-1-\tau}a_{i}b^{*}_{(i+\tau)}, & 0 \leq \tau < L,
\\
\sum_{i=0}^{L+\tau-1} a_{(i-\tau)}b^{*}_{i}, & -L< \tau < 0.  
\end{cases}
\end{equation}
The function $\mathcal{A}(\mathbf{a}, \mathbf{b})$ is referred to as aperiodic cross-correlation function (ACCF) of $\mathbf{a}$ and $\mathbf{b}$. If $\mathbf{a}=\mathbf{b}$ then this function is called aperiodic auto-correlation function (AACF) and denoted as $\mathcal{A}(\mathbf{a})$.
\par Further, the periodic cross-correlation function (PCCF) of $\mathbf{a}$ and $\mathbf{b}$ is defined as
\begin{equation}
\mathcal{P}(\mathbf{a}, \mathbf{b})(\tau)=
\begin{cases}
\sum_{i=0}^{L-1}a_{i}b^{*}_{(i+\tau)\mod L}, & 0 \leq \tau < L,
\\
\sum_{i=0}^{L-1} a_{i}b^{*}_{(i-\tau)\mod L}, & -L< \tau < 0.  
\end{cases}
\end{equation}
When $\mathbf{a}=\mathbf{b}$, then this function is called periodic auto-correlation function (PACF) and denoted as $\mathcal{P}(\mathbf{a})$ \cite{Adikary2016optimal}.
 \begin{definition}
Let $\mathbf{C}=\{\mathbf{C}_0,\mathbf{C}_1, \hdots ,\mathbf{C}_{K-1}\}$ be a collection of $K$ matrices (codes) of order $M\times L$. Define
\begin{align}
\mathbf{C_{\xi}}=
\begin{bmatrix}
\mathbf{a}_{0}^{\xi}\\
\mathbf{a}_{1}^{\xi}\\
\vdots\\
\mathbf{a}_{M-1}^{\xi}
\end{bmatrix},
\end{align}\\
where $\mathbf{a}_\upsilon^\xi$ ($0\leq \upsilon \leq M-1,0 \leq \xi \leq K-1$) is the $\upsilon$th row sequence or $\upsilon$th constituent sequence. Then the ACCF of two codes $\mathbf{C_{\xi_1}},\mathbf{C_{\xi_2}}\in \textbf{C}$ is defined as
\begin{equation}
    \gamma(\mathbf{C_{\xi_1}},\mathbf{C_{\xi_2}})(u)=\sum_{\upsilon=0}^{M-1}{\gamma(\mathbf{a}_{\upsilon}^{\xi_1}, \mathbf{a}_{\upsilon}^{\xi_2})(u)}.
\end{equation}
\end{definition}
\begin{definition}
  Let \textbf{C} be a code set satisfying the following correlation properties
   \begin{equation}
       \gamma(\mathbf{C_{\xi_1}},\mathbf{C_{\xi_2}})(u)=
      \begin{cases}
        LM, & \xi_1=\xi_2\ \text{and}\ u=0,  \\
        0,  & \xi_1=\xi_2\ \text{and}\ 0<|u|<L, \\
        0,  & \xi_1\neq \xi_2\ \text{and}\ |u|<L.
      \end{cases}
    \end{equation}  
  Then $\textbf{C}$ is known as $(K,M,L)$-MOGCS. If $K=M$, it is referred to be CCC set and we write it as $(K,K,L)$-CCC. Additionally, each code of $\textbf{C}$ is known as GCS \cite{SaMaLi}. When $M=2$ it is called GCP \cite{kumar2021direct}.\label{def2}
\end{definition}
\begin{definition}
  Let $\boldsymbol{\mathcal{Z}}=\{\mathbf{z}_0,\mathbf{z}_1,\hdots,\mathbf{z}_{K-1}\}$ be a set of $K$ sequences each of length $L$, i.e.,
  \[\mathbf{z}_i=(\mathbf{z}_{i0},\mathbf{z}_{i1},\hdots,\mathbf{z}_{iL-1}),\ \ 0\leq i\leq K-1.\]
  Then, $\boldsymbol{\mathcal{Z}}$ is referred to be $(K,L,Z)$-ZCZ sequence set for $0\leq i,j\leq K-1$, if $\boldsymbol{\mathcal{Z}}$ satisfies the following,
  \begin{equation}
      \mathcal{P}(\mathbf{z}_i,\mathbf{z}_j)(u)=
      \begin{cases}
       0, & i=j\ \text{and}\ 1\leq |u| < Z,\\
       0, & i\neq j \ \text{and}\ 0\leq |u| < Z,\\
       L, & i=j\ \text{and}\ u=0,
       \end{cases}
  \end{equation}
  where $Z$ is termed as ZCZ width.
\end{definition} 

\begin{definition}\emph{(Tang-Fan-Matsufuji Bound \cite{TaFaMa})}
Let $\boldsymbol{\mathcal{Z}}$ be any ZCZ sequence set with parameter $(K,L,Z)$. Then, $KZ\leq L$ and if $KZ=L$, then $\boldsymbol{\mathcal{Z}}$ is said to be optimal. If for the larger value of $K$, $KZ\approx L$, $\boldsymbol{\mathcal{Z}}$ is said to be asymptotically optimal. For the binary case, it is widely accepted that the bound is reduced to $2KZ\leq L$ \cite{matsu}. 

\end{definition}
\begin{definition}
Let $\boldsymbol{\mathcal{Z}}$ be a $(K,L,Z)$-ZCZ sequence set then $\boldsymbol{\mathcal{Z}}$ is called $(K,L,Z)$-Golay-ZCZ sequence set if it additionally satisfies the GCS properties.
\end{definition}
\begin{definition}
  Let $\mathbf{C}=\{\mathbf{C}_0,\mathbf{C}_1, \hdots ,\mathbf{C}_{K-1}\}$ be a $(K,K,L)$-CCC. Then $\mathbf{C}$ is called $(K,K,L,Z)$-CC-ZCZ code set if each $\mathbf{C}_i,~\forall~ i=0,1,\hdots,k-1$ is $(K,L,Z)$-Golay-ZCZ sequence set. 
\end{definition}
\setlength{\jot}{10pt}
\subsection{Multivariable Functions and Corresponding Sequences \cite{SaLiMa}}
Let $\mathbf{x}=(x_1,x_2,\hdots,x_{k-1})\in \mathbb{Z}_p^k$, be a vector of finite length, where $p$ is a prime number and $k$ is a positive integer. It is clear that $x_i\in \mathbb{Z}_p$ for $1\leq i\leq k.$ Now, we take a specific collection of monomials of degree at most $r$ over the variables $x_1,x_2,\hdots,x_k$ as follows
\begin{equation}
  \mathscr{M}(\mathbf{x},r)=\{x_{1}^{j_1}x_{2}^{j_2}\hdots x_{k}^{j_{k}}: j_i\in \{0,1\} \ \text{for}\ 1\leq i\leq k,j_1+j_2+\cdots+j_{k}\leq r\},
\end{equation}
where $0\leq r\leq k$. A monomial's degree in $\mathscr{M}$ is determined by the number of independent variables associated with it and this can be verified by examining the expressions of the monomials in $\mathscr{M}$.
A linear combination of monomials in $\mathscr{M}$ with $\mathbb{Z}_q$-valued coefficients, where $q$ is a positive integer, leads to a multivariable function $f(x_1,x_2,\hdots,x_{k}):\mathbb{Z}_p^k\rightarrow\mathbb{Z}_q$ which has maximum degree $r$. Let $\mathscr{F}(\mathscr{M}(\mathbf{x},r),q)$ be the set of all multivariable functions $f(x_1,x_2,\hdots,x_{k}):\mathbb{Z}_p^k\rightarrow\mathbb{Z}_q$, over the monomials in $\mathscr{M}$ and defined as
\begin{equation}
    \mathscr{F}(\mathscr{M}(\mathbf{x},r),q)=\bigg\{\sum_{(j_1,j_2,\hdots,j_k)\in \{0,1\}^k}{c_{j_1,j_2,\hdots,j_k}x_{1}^{j_1}x_{2}^{j_2}\hdots x_{k}^{j_{k}}}:c_{j_1,j_2,\hdots,j_k}\in \mathbb{Z}_q\bigg\},
\end{equation}
where $r$ is the maximum degree of a multivariable function $f$ in $\mathscr{F}(\mathscr{M}(\mathbf{x},r),q)$. When $p=2$, the set of multivariable functions $\mathscr{F}(\mathscr{M}(\mathbf{x},r),q)$ is reduced to a set of $\mathbb{Z}_q$-valued
GBFs of maximum degree $r$ over the vector variable $\mathbf{x}$ .
\par For a multivariable function, $f\in \mathscr{F}(\mathscr{M}(\mathbf{x},r),q)$, the $\mathbb{Z}_q$-valued sequence $\theta(f)$ and the complex-valued sequence $\Theta(f)$ having length $p^k$ are defined as
\begin{equation}
    \theta(f)=({f_0}, {f_1}, \hdots,{f_{p^k-1}}),\label{eq:11}
\end{equation}
\begin{equation}
    \Theta(f)=(\omega^{f_0}, \omega^{f_1}, \hdots, \omega^{f_{p^k-1}}).
\end{equation}
where $f_i=f(i_0,i_1,\hdots,i_{k-1})$, $\omega=\exp(2\pi\sqrt{-1}/q)$, and $(i_0,i_1,\hdots,i_{k-1})$ is the 
$p$-ary vector representation of $i$.
\subsection{Generalized Reed-Muller (GRM) Code}
For a prime $p$, let $\lambda=p^n$ and $n\geq 1$, then codewords of $\lambda$-ary GRM codes consist of the sequence corresponding to multivariable polynomials
over the field $\mathbb{F}_\lambda$, where $\lambda=p^n$ and $n\geq 1$. Let us consider a polynomial ring $\mathbb{F}_\lambda[x_1,x_2,\hdots,x_m]$ with $m$ variables. The GRM code with parameters $m$ and $r$ consists of all the sequences corresponding to polynomials with $m$ variables and degree no larger than $r$.
\begin{definition}
The $r$-th order $r\leq \lambda$, $\lambda$-ary GRM code denoted by $GRM_\lambda(m, r)$, is defined by the set of $\lambda$-ary vectors as
\begin{equation}
GRM_\lambda(m,r)=\{\theta(f):f\in\mathbb{F}_\lambda[x_1,x_2,\hdots,x_m],~deg(f)\leq r\} .
\end{equation}
\end{definition}It is to be noted that we always have $x^\lambda=1$ in $\mathbb{F}_\lambda$, so we only need to consider polynomials in which the degree of each $x_i$ is no larger than $\lambda-1$. All such polynomials with degree no larger than $r$ are linear combinations of the following set of monomials
\begin{equation}
    \mathcal{N}(r)=\Big\{x_1^{j_1}x_2^{j_2},\hdots x_m^{j_m}:j_i=0,1,2,\hdots,\lambda-1~\text{and}~\sum_{i=1}^{m}{j_i}\leq r\Big\}.
\end{equation}
Using combinatorics, we can always prove that the number of monomials in the set $\mathcal{N}(r)$ is, $|\mathcal{N}(r)|=\sum_{d=0}^{r}{\binom{m-1+d}{d}}$. It is well known that $\mathbb{F}_\lambda[x_1,x_2,\hdots,x_m]$ forms a vector space over the field $\mathbb{F}_\lambda$, further it can be easily proved that $GRM_\lambda(m,r)$ is its subspace. Therefore, $GRM_\lambda(m,r)$ is a linear code with code length $n=\lambda^m$ and code dimension $\sum_{d=0}^{r}{{m-1+d\choose d}}$. Moreover, if we arrange sequences corresponding to monomials in $\mathcal{N}(r)$ as the rows of a matrix then this matrix forms a generator matrix of $GRM_\lambda(m,r)$.
\begin{lemma}[\cite{KaSh}]
   For $\lambda=p^n$ and $n\geq 1$, minimum Hamming distance of $GRM_{\lambda}(m,r)$ is $(R+1).p^Q,$ where $R$ is the remainder and $Q$ is the quotient resulting from dividing $m(p-1)-r$ by $p-1$.\label{l0}  
\end{lemma}
\subsection{Peak-to-Mean Envelope Power Ratio (PMEPR)}
In spite of several benefits arising from the OFDM technique, its wide acceptance has been hindered by its high PMEPR in uncoded signals. In this subsection, we define PMEPR for OFDM signals.
\par
One can model the OFDM signal for a complex valued word $\mathbf{B}=(B_1,B_2,\hdots,B_N)$ of length $N$ as the real part of the
\begin{equation}
    \mathcal{S}_\mathbf{B}(\tau)=\sum_{i=1}^{N}{B_ie^{2\pi(i+\zeta)(\tau)\sqrt{-1}}},
\end{equation}
where $0\leq\tau<1$ and $\zeta$  is a positive constant. The sequence $\mathbf{B}=(B_1,B_2,\hdots,B_N)$ is termed as modulating codeword of the OFDM symbol. The PMEPR of word $\mathbf{B}$ is defined as
\begin{equation}
    PMEPR(\mathbf{B})= \frac{1}{N}\sup_{0\leq\tau<1}{|\mathcal{S}_{\mathbf{B}}(\tau)|^2}.
\end{equation}
It is to be observed that the largest value that PMEPR can have is $N$. 
However, it is preferable to use codewords with a lower PMEPR than $N$. Additionally, the PMEPR of GCS of size $M$ is bounded above by $M$ \cite{Paterson}.

\section{Proposed Construction of CC-ZCZ Code Set}
\begin{figure}[]
     \centering
         \includegraphics[trim={0 0 19cm 0},width=15cm,height=10cm]{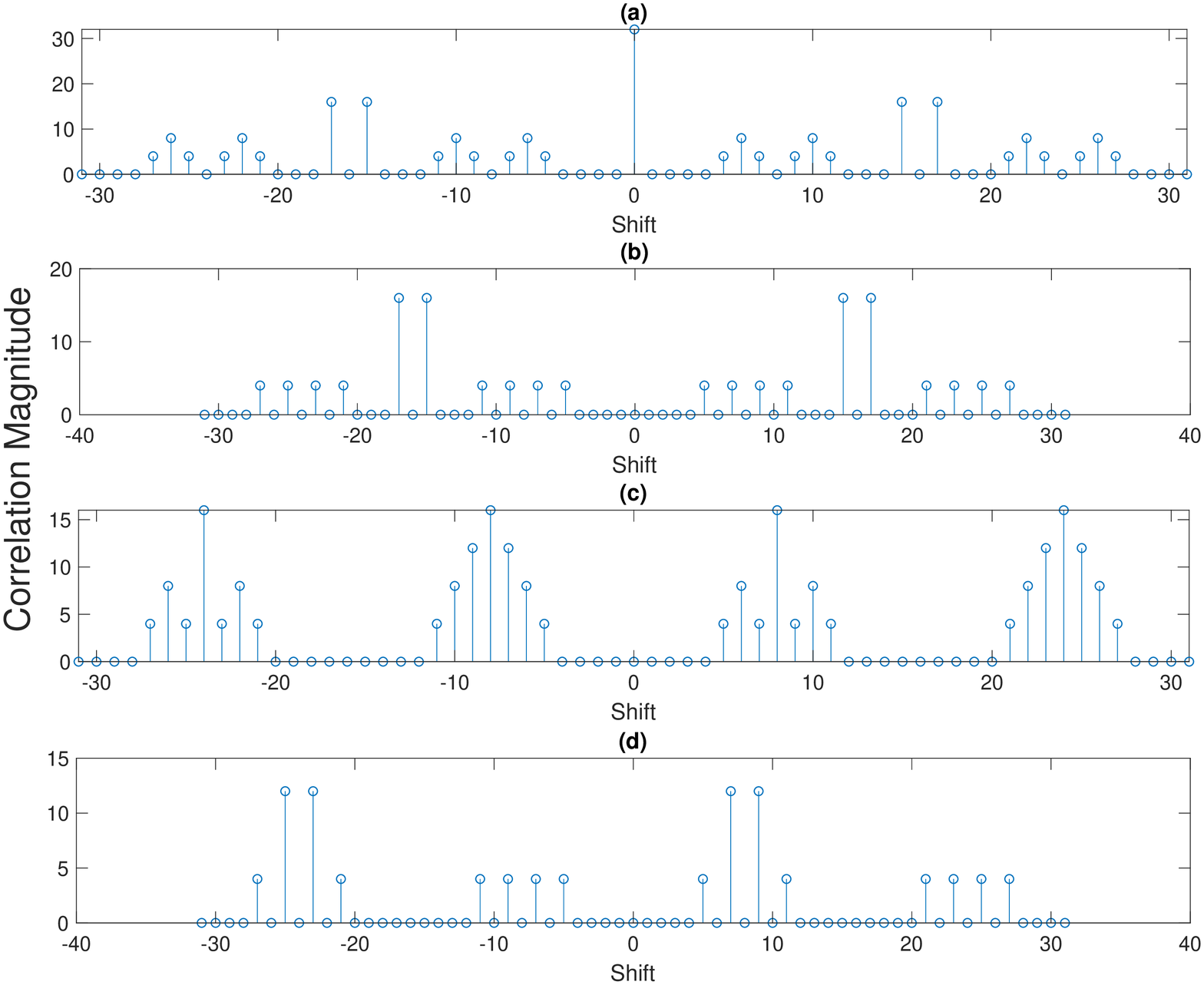}
        \caption{$\mathbf{(a)}$ Auto-correlation plot for \{$\theta(a_\mathbf{0}^\mathbf{0})$, $\theta(a_\mathbf{0}^\mathbf{1})$, $\theta(a_\mathbf{0}^\mathbf{2})$, $\theta(a_\mathbf{0}^\mathbf{3})$\}, $\mathbf{(b)}$ Cross-correlation plot for $\{(\theta(a_\mathbf{0}^\mathbf{1}),\theta(a_\mathbf{0}^\mathbf{2})),(\theta(a_\mathbf{0}^\mathbf{3}),\theta(a_\mathbf{0}^\mathbf{4}))\}$, $\mathbf{(c)}$ Cross-correlation plot for $\{(\theta(a_\mathbf{0}^\mathbf{1}),\theta(a_\mathbf{0}^\mathbf{3})),(\theta(a_\mathbf{0}^\mathbf{2}),\theta(a_\mathbf{0}^\mathbf{4}))\}$ $\mathbf{(d)}$ Cross-correlation plot for $\{(\theta(a_\mathbf{0}^\mathbf{1}),\theta(a_\mathbf{0}^\mathbf{4})),\allowbreak(\theta(a_\mathbf{0}^\mathbf{2}),\theta(a_\mathbf{0}^\mathbf{3}))\}.$}
        \label{F1}
\end{figure}

In this section, we provide a construction of CC-ZCZ code set of prime-power length. For that we start with a result which provides the construction of CCC of prime-power length \cite{SaLiMa}. For the ease of representation, we present this result in a modified way without loosing the essence of the result.
\begin{lemma}[\cite{SaLiMa}]\label{l1} 
  For an integer $m\geq2,$ let $\{1,2,\hdots,m\}$ is divided into $k$ partitions namely, $E_1,E_2,\hdots,E_k$ where $1\leq k\leq m-1$. Further, let us also assume that $\lambda=p^n$ and $n\geq1$ and $\pi_{\beta}$ be bijection mapping from $\{1,2,\hdots,n_{\beta}\}$ to $E_{\beta}$ where $n_{\beta}=|E_{\beta}|\geq 1,~ \forall{\beta=2,3,\hdots,k}$ and $n_1=|E_1|\geq2$. Now, we define a multivariable function $f:\mathbb{Z}_p^m\rightarrow\mathbb{Z}_\lambda$ as
  \begin{equation}
      f(x_1,x_2,\hdots,x_m)=\frac{\lambda}{p}\sum_{\beta=1}^{k}{\sum_{\gamma=1}^{n_{\beta}-1}{x_{\pi_{\beta}(\gamma)}x_{\pi_{\beta}(\gamma+1)}}}+\sum_{\alpha=1}^{m}{g_{\alpha}x_{\alpha}},\label{eq:12}
  \end{equation}
  where $g_{\alpha}\in \mathbb{Z}_\lambda$ for $\alpha=1,2,\hdots, m$. Further, let $\mathbf{u_k}=[u_1,u_2,\hdots,u_k],~\mathbf{v_k}=[v_1,v_2,\hdots,v_k]$ be $p$-ary representation for $0\leq \mathbf{u_k,v_k}\leq p^k-1$. Define
  \begin{equation}
      a_{\mathbf{u_k}}^{\mathbf{v_k}}=a_{u_1}^{v_1}(E_1)+a_{u_2}^{v_2}(E_2)+\cdots+a_{u_k}^{v_k}(E_k),\label{eq:13}
  \end{equation}
  where for any $1\leq \beta \leq k$, multivariable function $a_{u_\beta}^{v_\beta}(E_\beta)$ is defined as
  \begin{equation}
      a_{u_\beta}^{v_\beta}(E_\beta)=\frac{\lambda}{p}\sum_{\gamma=1}^{n_{\beta}-1}{x_{\pi_{\beta}(\gamma)}x_{\pi_{\beta}(\gamma+1)}}+\sum_{\gamma=1}^{n_\beta}{g_{\pi_\beta{(\gamma)}}x_{\pi_\beta{(\gamma)}}}+\frac{\lambda}{p}x_{\pi_{\beta}(1)}u_\beta+\frac{\lambda}{p}x_{\pi_{\beta}(n_\beta)}v_\beta.\label{eq:14}
  \end{equation}
  Let us also define
  \begin{equation}
      \mathbf{C_{u_k}}=[\Theta( a_{\mathbf{u_k}}^{\mathbf{v_k}})]_{0\leq \mathbf{v_k}\leq p^k-1}=
      \begin{bmatrix}
        \Theta( a_{\mathbf{u_k}}^0)\\
        \Theta( a_{\mathbf{u_k}}^1)\\
        \vdots\\
        \Theta( a_{\mathbf{u_k}}^{p^k-1})
      \end{bmatrix}.\label{eq:15}
  \end{equation}
  Then $\mathbf{C}=\{\mathbf{C_{u_k}}:0\leq \mathbf{u_k}\leq p^k-1\}$ is $(p^k,p^k,p^m)$-CCC.
  \end{lemma}
  \begin{theorem}\label{th1}
       For a fixed $\mathbf{u_k}$, $\mathbf{C_{u_k}}$ is GCS as defined in \emph{Lemma \ref{l1}}. Let $\mathbf{C'_{u_k}}$ be $\mathbf{C_{u_k}}$ together with the permutation $\pi_{\beta}(1)=m-\beta+1$ for $\beta=1,2,\hdots,k$. Then $\mathbf{C'_{u_k}}$ is $(p^k,p^m,(p-1)p^{\pi_1(2)-1})$-Golay-ZCZ sequence set and hence $\mathbf{C'}=\{\mathbf{C'_{u_k}}:0\leq \mathbf{u_k}\leq p^k-1\}$ is $(p^k,p^k,p^m,(p-1)p^{\pi_1(2)-1})$-CC-ZCZ code set.  
  \end{theorem}
  \begin{IEEEproof}
      Please see Appendix A.
  \end{IEEEproof}
  \begin{remark}
  If we have $\pi_1(2)=m-k$, then from \emph{Theorem \ref{th1}}, $\mathbf{C'_{u_k}}$, is $(p^k,p^m,(p-1)p^{m-k-1})$-ZCZ $ \forall~0\leq \mathbf{u_k}\leq p^k-1$, which implies $KZ=(p-1)p^{m-1}$ and $L=p^m$. Now, for $\lambda=2$, $KZ=2L$ and hence $\mathbf{C'_{u_k}}$ is optimal ZCZ sequence set. But if $p\neq2$ then for larger value of $p$, $(p-1)\approx p$ and hence $KZ\approx L$, therefore, in this case, $\mathbf{C'_{u_k}}$ is asymptotically optimal.
  \end{remark}
  According to \emph{Theorem \ref{th1}}, we can provide a construction of $(p^k,p^k,p^m,(p-1)p^{\pi_1(2)-1})$-CC-ZCZ code sets having variable ZCZ width $\pi_{1}(2)$. It is desirable to have a larger ZCZ width, we take $\pi_{1}(2)=m-k$ to propose a family of $(p^k,p^k,p^m,(p-1)p^{m-k-1})$-CC-ZCZ code sets having larger ZCZ width. 
  Furthermore, each code of CC-ZCZ is $(p^k,p^m,(p-1)p^{m-k-1})$-Golay-ZCZ sequence set and it is optimal \emph{(Tang-Fan-Matsufuji Bound)} for $\lambda=2$ and asymptotically optimal for $\lambda\neq2$. To support our construction, we present examples below.
\begin{example}(\emph{Binary Case})
Taking $\lambda = 2,p=2, k = 2,$ and $m = 5$, we let
$E_1 = \{1, 3, 5\}$ and $E_2 = \{2, 4\}$ be a partition of $\{1, 2, 3, 4, 5\}$. Also let $\pi_1(1) = 5, \pi_1(2) = 3, \pi_1(3) = 1, \pi_2(1) = 4$ and $\pi_2(2) = 2$. Let $f:\mathbb{Z}_2^5\rightarrow\mathbb{Z}_2$ be defined as
\[f(x_1,x_2,\hdots,x_5)=x_5x_3+x_3x_1+x_4x_2+x_1+x_3.\]
Further, let $\mathbf{u_k}=0$, Then
\begin{align}
\mathbf{C'_0}=
\begin{bmatrix}
      \theta(a_\mathbf{0}^\mathbf{0})\\
      \theta(a_\mathbf{0}^\mathbf{1})\\
      \theta(a_\mathbf{0}^\mathbf{2})\\
      \theta(a_\mathbf{0}^\mathbf{3})
    \end{bmatrix}
    &=
    \begin{bmatrix}
     \theta(f+x_5.0+x_4.0+x_1.0+x_2.0)\\
      \theta(f+x_5.0+x_4.0+x_1.1+x_2.0)\\
      \theta(f+x_5.0+x_4.0+x_1.0+x_2.1)\\
      \theta(f+x_5.0+x_4.0+x_1.1+x_2.1)
    \end{bmatrix}\nonumber
    \end{align}
    \begin{align}
    &=
\begin{bmatrix}
0 1 0 1 1 1 1 1 0 1 1 0 1 1 0 00 1 0 1 0 0 0 0 0 1 1 0 0 0 1 1\\
0 0 0 0 1 0 1 0 0 0 1 1 1 0 0 1 0 0 0 0 0 1 0 1 0 0 1 1 0 1 1 0\\
0 1 1 0 1 1 0 0 0 1 0 1 1 1 1 1 0 1 1 0 0 0 1 1 0 1 0 1 0 0 0 0\\
0 0 1 1 1 0 0 1 0 0 0 0 1 0 1 0 0 0 1 1 0 1 1 0 0 0 0 0 0 1 0 1
\end{bmatrix},\nonumber
\end{align}
where the above matrix is phase matrix and $\mathbf{C'_{0}}$ is a $(4,32,4)$-Golay-ZCZ sequence set. Moreover, we can also derive $\mathbf{C'_{1}},\mathbf{C'_{2}}$ and $\mathbf{C'_{3}}$ and these are also $(4,32,4)$-Golay-ZCZ sequence sets. Hence $\mathbf{C'}=\{\mathbf{C'_0},\mathbf{C'_1},\mathbf{C'_2},\mathbf{C'_3}\}$ is a $(4,4,32,4)$-CC-ZCZ code set. Since $K=4,~ L=32,~ Z=4$ which implies $KZ=16$. Hence, $2KZ=L$. Therefore, $\mathbf{C'}_i$ is optimal by \emph{Tang-Fan-Matsufuji} Bound. A graphical representation of periodic correlation of sequences in $\mathbf{C'_0}$ is shown in Fig. \ref{F1} 
  \end{example}
  
  \begin{example}(\emph{Non-binary case}) Here let $\lambda=3,k=2,m=3$ and $p=3$. Also let $E_1=\{2,3\},~E_2=\{1\}$ be a partition of $\{1,2,3\}$. Further, let $\pi_1(1)=3,\pi_1(2)=2,$ and $\pi_2(1)=1$. Define $f:\mathbb{Z}_3^3\rightarrow\mathbb{Z}_3$
  as $f(x_1,x_2,x_3)=x_3x_2$. Now, let us fix $\mathbf{u_k=0}$ then
  \begin{align}
      \mathbf{C'_0}&=
      \begin{bmatrix}
      a_\mathbf{0}^\mathbf{0}\\
      a_\mathbf{0}^\mathbf{1}\\
      a_\mathbf{0}^\mathbf{2}\\
      a_\mathbf{0}^\mathbf{3}\\
      a_\mathbf{0}^\mathbf{4}\\
      a_\mathbf{0}^\mathbf{5}\\
      a_\mathbf{0}^\mathbf{6}\\
      a_\mathbf{0}^\mathbf{7}\\
      a_\mathbf{0}^\mathbf{8}
\end{bmatrix}
 =
    \begin{bmatrix}
      f+x_3.0+x_1.0+x_2.0+x_1.0\\
      f+x_3.0+x_1.0+x_2.1+x_1.0\\
      f+x_3.0+x_1.0+x_2.2+x_1.0\\
      f+x_3.0+x_1.0+x_2.0+x_1.1\\
      f+x_3.0+x_1.0+x_2.1+x_1.1\\
      f+x_3.0+x_1.0+x_2.2+x_1.1\\
      f+x_3.0+x_1.0+x_2.0+x_1.2\\
      f+x_3.0+x_1.0+x_2.1+x_1.2\\
      f+x_3.0+x_1.0+x_2.2+x_1.2
    \end{bmatrix}
    =
\begin{bmatrix}
0 0 0 0 0 0 0 0 0 0 1 2 0 1 2 0 1 2 0 2 1 0 2 1 0 2 1\\
0 1 2 0 1 2 0 1 2 0 2 1 0 2 1 0 2 1 0 0 0 0 0 0 0 0 0\\
0 2 1 0 2 1 0 2 1 0 0 0 0 0 0 0 0 0 0 1 2 0 1 2 0 1 2\\
0 0 0 1 1 1 2 2 2 0 1 2 1 2 0 2 0 1 0 2 1 1 0 2 2 1 0\\
0 1 2 1 2 0 2 0 1 0 2 1 1 0 2 2 1 0 0 0 0 1 1 1 2 2 2\\
0 2 1 1 0 2 2 1 0 0 0 0 1 1 1 2 2 2 0 1 2 1 2 0 2 0 1\\
0 0 0 2 2 2 1 1 1 0 1 2 2 0 1 1 2 0 0 2 1 2 1 0 1 0 2\\
0 1 2 2 0 1 1 2 0 0 2 1 2 1 0 1 0 2 0 0 0 2 2 2 1 1 1\\
0 2 1 2 1 0 1 0 2 0 0 0 2 2 2 1 1 1 0 1 2 2 0 1 1 2 0
\end{bmatrix},
  \end{align}
   $\mathbf{C'_0}$ is a $(9,27,2)$-Golay-ZCZ sequence set. Similarly, $\mathbf{C'_{u_k}}$ for $1\leq\mathbf{u_k}\leq8$ are also $(9,27,2)$-Golay-ZCZ sequence set. Hence $\mathbf{C'}=\{\mathbf{C'_0},\mathbf{C'_1},\hdots,\mathbf{C'_8}\}$ is a $(9,9,27,2)$-CC-ZCZ code set.
  \end{example}
  \section{CC-ZCZ Code Sets from GRM Codes}
 In this section, a relation between CC-ZCZ code set and elements of second-order coset of $GRM_p(m,1)$ is established for $\lambda=p$. The precise number of cosets and number of CC-ZCZ code set in each coset is also given. Hamming distance of Golay-ZCZ sequence is also calculated.
 \par
It can be obtained that if we have a sequence in $\mathbf{C'_{u_k}}$, then it is a codeword belonging to second-order coset $\mathcal{Q}_1 + GRM_p(m,1)$ where $\mathcal{Q}_1$ is given by
\begin{equation}
        \mathcal{Q}_1=\sum_{\beta=1}^{k}{\sum_{\gamma=1}^{n_{\beta}-1}{x_{\pi_{\beta}(\gamma)}x_{\pi_{\beta}(\gamma+1)}}}. \label{eq:100}
\end{equation}
Conversely, 
if any sequence is in the coset $\mathcal{Q}_1 + GRM_p(m,1)$ then it belongs to a certain $(p^k,p^m,(p-1)p^{\pi_{1}(2)-1})$-Golay-ZCZ sequence set. Now, let us consider the constructed $(p^k,p^m,(p-1)p^{m-k-1})$-Golay-ZCZ sequence family by taking $\pi_{1}(2) = m- k$.
\begin{corollary}
The number of coset representative $\mathcal{Q}_1$ in the second-order coset of the form $\mathcal{Q}_1+GRM_p(m,1)$
is given by 
\begin{equation}
    \sum_{n_1+n_2+\cdots+n_k=m}{{\binom{m-k-1}{n_1-2,n_2-1,\hdots,n_k-1}}(n_1-2)!\prod_{\beta=2}^{k}{(n_{\beta}-1)!}}.
\end{equation}\label{cor1}
\end{corollary}

\begin{IEEEproof}
 Since, we have $\pi_{1}(1)=m$ and $\pi_{1}(2)=m-k$, so we can get
$(n_1-2)!$ different quadratic forms of the type
\begin{equation}
  x_{\pi_{1}(1)}x_{\pi_{1}(2)} + x_{\pi_{1}(2)}x_{\pi_{1}(3)} + \cdots+ x_{\pi_{1}(n_1-1)}x_{\pi_{1}(n_1)},
\end{equation}
where $\pi_{1}(i)$ is a permutation of $n_1 - 2$ integers where $i = 3, 4,\hdots,n_1$. Similarly, there are $(n_\beta - 1)!$ different quadratic forms of the type $\sum_{\beta=1}^{n_{\beta}-1}{x_{\pi_{\beta}(\gamma)}x_{\pi_{\beta}(\gamma+1)}}$ for $\beta = 2, 3,\hdots,k$ as $\pi_{\beta}(1)$ is restricted to $m - \beta + 1$.
In addition, we have the condition $n_1 + n_2 + \cdots + n_k = m$
where $n_\beta = |E_\beta|$ and non-empty sets $E_1, E_2,\hdots,E_k$ form a partition of $\{1, 2,\hdots,m\}$. Therefore, we can explicitly determine the number of distinct coset representatives $\mathcal{Q}_1$ as
\begin{equation*}
    \sum_{n_1+n_2+\cdots+n_k=m}{{\binom{m-k-1}{n_1-2,n_2-1,\hdots,n_k-1}}(n_1-2)!\prod_{\beta=2}^{k}{(n_{\beta}-1)!}}.
\end{equation*}
\end{IEEEproof}

\begin{corollary}
Let $\mathcal{Q}_1+GRM_p(m,1)$ be any second-order coset
with coset representative $\mathcal{Q}_1$ 
and let $\pi_{1}(2) =m-k$. Then the coset $\mathcal{Q}_1+GRM_p(m,1)$ consists of $p^{m-k+1}$ and $p^{m+1}$ distinct $p$-ary $(p^k,p^k,p^m,(p-1)p^{m-k-1})$-CC-ZCZ code set and $(p^k,p^m,(p-1)p^{m-k-1})$-Golay-ZCZ sequence set respectively.\label{cor2}
\end{corollary}
\begin{IEEEproof}
    We have from \emph{Theorem \ref{th1}}, that a codeword $\mathbf{c}$ in $\mathcal{Q}_1+GRM_p(m,1)$ lies in the $(p^k,p^m,(p-1)p^{m-k-1})$-Golay-ZCZ sequence set 
\begin{equation}
       \mathbf{C'_{u_k}}=\bigg\{f+\sum_{\beta=0}^{k-1}{x_{\pi_{\beta+1}{(1)}}u_{\beta+1}}+\sum_{\beta=1}^{k}{x_{\pi_{\beta}{(n_\beta)}}v_{\beta}}+b_0:v_{\beta}\in\mathbb{Z}_p,~1\leq \beta\leq k\bigg\}
       \label{eq:50}.
\end{equation}
So it can easily be obtained that for a fixed value of $\mathbf{u_k}$, $\mathcal{Q}_1+GRM_p(m,1)$ consists of $p^{m-k+1}$ distinct $p$-ary $(p^k,p^m,(p-1)p^{m-k-1})$-Golay-ZCZ sequence set. Hence, we have $p^{m-k+1}$ distinct $p$-ary $(p^k,p^k,p^m,(p-1)p^{m-k-1})$-CC-ZCZ code set. Since $\mathbf{u_k}$ can vary in $p^k$ ways, therefore, $\mathcal{Q}_1+GRM_p(m,1)$ consists of $p^{m-k+1}\cdot p^k=p^{m+1}$ distinct $p$-ary $(p^k,p^m,(p-1)p^{m-k-1})$-Golay-ZCZ sequence set.
\end{IEEEproof}
\par Since the codes in our construction lie inside the second-order cosets of first-order $GRM$ codes, therefore, they have the high Hamming distance. Hamming distance of a code measures how efficient a code is to detect and correct errors. If the Hamming distance of a code is $h$ then we can correct the errors of Hamming weight less than $h/2$. Therefore, in the next corollary, we prove that minimum Hamming distance of the constructed CC-ZCZ code set is equal to minimum Hamming distance of $GRM_p(m,1)$.
\begin{corollary}
  The minimum Hamming distance of $(p^k,p^m,(p-1)p^{m-k-1})$-Golay-ZCZ sequence sets which are constructed in \emph{Theorem \ref{th1}} is $(p-1)p^{m-1}$.\label{cor3}
\end{corollary}
 \begin{IEEEproof}
Since, $\mathbf{C'_{u_k}}$ is contained in the second-order coset 
\begin{equation}
    \mathcal{Q}_1 + GRM_p(m,1)=\{\mathcal{Q}_1 + c : c\in GRM_p(m,1)\}.
\end{equation}
Hence, according to \emph{Lemma \ref{l0}}, $\mathbf{C'_{u_k}}$ has the minimum Hamming distance $(p-1)p^{m-1}$ . 
\end{IEEEproof} 
\par
Considering $p$-ary sequences, let $N(\mathcal{Q}_1)$ be the number of coset representatives $\mathcal{Q}_1$, $N(GZ)$ be the number of distinct $(p^k,p^m,(p-1)p^{m-k-1})$-Golay-ZCZ sequence set in a coset $\mathcal{Q}_1+GRM_p(m,1)$, $N(CZ)$ be the number of distinct $(p^k,p^k,p^m,(p-1)p^{m-k-1})$-CC-ZCZ code set in a coset $\mathcal{Q}_1+GRM_p(m,1)$, and $d(\mathcal{Q}_1+GRM_p(m,1))$ be the minimum
Hamming distance of $\mathcal{Q}_1+GRM_p(m,1)$. 
Taking $p=3$ in \emph{corollary} \ref{cor1}, \emph{corollary} \ref{cor2}, and \emph{corollary} \ref{cor3}, Table \ref{t1} shows the values of $N(\mathcal{Q}_1),N(GZ),N(CZ)$ and $d(\mathcal{Q}_1 + GRM_3(m,1))$ for lengths $27$, $81$, and $243$. 
Taking $m = 5$ and $k = 1$, it can be observed from Table \ref{t1} that there are $6$ second-order cosets $\mathcal{Q}_1 + GRM_3(5,1)$ and each coset consists of $243$ distinct $(3,3,243,54)$-CC-ZCZ code set. If we want to increase the set size from $3$ to $9$, we need to take $k = 2$. Then, Table \ref{t1} gives that we can obtain $81$, $(9,9,243,18)$-CC-ZCZ code set with reduced ZCZ width from one of $6$ different second-order cosets $\mathcal{Q}_1 + GRM_3(5,2)$.
\begin{table}{}
\caption{Calculated Values of $N(\mathcal{Q}_1)$, $N(GZ)$, $(K,L,Z)$ and $d(\mathcal{Q}_1+GRM_3(m,1))$.}
\resizebox{\textwidth}{!}{
\begin{threeparttable}
    \begin{tabular}{|c|c|c|c|c|c|c|}
    \hline
       $\mathbf{m}$ & $\mathbf{k}$ & $\mathbf{(K,K,L,Z)}$ & $\mathbf{N(\mathcal{Q}_1)}$  & $\mathbf{N(CZ)}$ & $\mathbf{N(GZ)}$ & $\mathbf{d(\mathcal{Q}_1+GRM_3(m,1))}$\\ \hline
       \multirow{2}{*}{$3$} & $1$ & $(3,3,27,6)$ & $1$ & $27$ & \multirow{2}{*}{$81$} & $18$\\
       & $2$ & $(9,9,27,2)$ & $1$ & $9$ & & $18$\\
       \hline
       \multirow{3}{*}{$4$} & $1$ & $(3,3,81,18)$ & $2$ & $81$ & \multirow{3}{*}{$243$} & $54$\\
       & $2$ & $(9,9,81,6)$ & $2$ & $27$ & & $54$\\
       & $3$ & $(27,27,81,2)$ & $1$ & $9$ &  & $54$\\
       \hline
       \multirow{4}{*}{$5$} & $1$ & $(3,3,243,54)$ & $6$ & $243$ & \multirow{4}{*}{$729$} & $162$\\
       & $2$ & $(9,9,243,18)$ & $6$ & $81$ & & $162$\\
       & $3$ & $(27,27,243,6)$ & $3$ & $27$ & & $162$\\
       & $4$ & $(81,81,243,2)$ & $1$ & $9$ &  & $162$\\
       \hline
    \end{tabular}
\end{threeparttable}\label{t1}
}
\end{table}
  \section{PMEPR of Proposed CC-ZCZ Code Set}
  In this section, a method has been discussed to reduce column sequence PMEPR of proposed CC-ZCZ code set. Since in the proposed CC-ZCZ code set, each code is Golay-ZCZ sequence set of size $p^k$ and hence its row sequence PMEPR is bounded by $p^k$. But column sequence PMEPR of CC-ZCZ can further be reduced. 
If we want to generate $j$th column of $\mathbf{C'_{u_k}}$, $0\leq j\leq p^m-1$ given in \eqref{eq:50} then only variable terms in \eqref{eq:50} is $\sum_{\beta=1}^{k}{x_{\pi_{\beta}{(n_\beta)}}v_{\beta}}$. Now, we add some constant to it and we can get, 
  \begin{align}
      \phi_j(v)&=\frac{\lambda}{p}\sum_{\beta=1}^{k-1}{v_{\pi'(\beta)}v_{\pi'(\beta+1)}}+\frac{\lambda}{p}\sum_{\beta=1}^{k}{j_{\pi_{\beta}{(n_\beta)}}v_{\beta}}+\frac{\lambda}{p}v_{\pi'(1)}t+\frac{\lambda}{p}v_{\pi'(k)}l
  \end{align}
 where $l,t,v\in \mathbb{Z}_p$, $\pi'$ be the permutation of symbols $1,2,\hdots,k$, $[v_1,v_2,\hdots,v_k]$ be the $p$-ary representation of $v$. By putting $m=k$ and $k=1$ in \emph{Lemma \ref{l1}} it can easily be seen that $\Theta(\phi_j(v))$ is a member of a GCS which has $p$ sequences each of length $p^k$. Hence its column sequence PMEPR is bounded by $p$. Now, let us redefine $\mathbf{C'_{u_k}}$ as
 \begin{equation}
       \mathbf{C'_{u_k}}=\bigg\{f+\frac{\lambda}{p}\sum_{\beta=0}^{k-1}{x_{\pi_{\beta+1}{(1)}}u_{\beta+1}}+\phi_x(v)+b_0:v_{\beta}\in\mathbb{Z}_p,~1\leq \beta\leq k\bigg\},
       \label{eq:52}
\end{equation}
where 
 \begin{equation}
      \phi_x(v)=\frac{\lambda}{p}\sum_{\beta=1}^{k-1}{v_{\pi'(\beta)}v_{\pi'(\beta+1)}}+\sum_{\beta=1}^{k}{x_{\pi_{\beta}{(n_\beta)}}v_{\beta}}+\frac{\lambda}{p}v_{\pi'(1)}t+\frac{\lambda}{p}v_{\pi'(k)}l.
\end{equation}
For some multivariable functions $g_1,g_2$ and constants $c_1,c_2$, it can easily be verified that
\begin{align}\nonumber
   \mathcal{A}(g_1+c_1)&=\mathcal{A}(g_1),\\ \nonumber
   \mathcal{A}(g_1+c_1,g_2+c_2)&=\mathcal{A}(g_1,g_2)\omega^{c_1-c_2},\label{eq:53}\\ 
   \mathcal{P}(g_1+c_1)&=\mathcal{P}(g_1), \\ \nonumber
   \mathcal{P}(g_1+c_1,g_2+c_2)&=\mathcal{P}(g_1,g_2)\omega^{c_1-c_2}.  \nonumber
\end{align}
Using \eqref{eq:53}, we can assure that redefining of $\mathbf{C'_{u_k}}$ as in \eqref{eq:52} affect its correlation properties by a multiplication of constant, i.e., $\mathbf{C'_{u_k}}$ appearing in \eqref{eq:52} is still a Golay-ZCZ sequence set 
whose column sequence PMEPR is bounded by $p$.
\begin{remark}
In \cite{ChWu}, the PMEPR of Golay-ZCZ sequence set is upper bounded by the number of sub-carriers, i.e, $2^k$. The PMEPR of Golay-ZCZ sequence set increases with the increase in the value of $k$. On the other hand, in the proposed Golay-ZCZ sequence set, the number of sub-carriers is $p^k$ and upper bound for its column sequence PMEPR is $p$. For $p=2$, we have $2^k$ sub-carriers and maximum column sequence PMEPR $2$, which remains same with increase in the value of $k$. So, the proposed construction have advantage over the construction presented in \cite{ChWu} as maximum column sequence PMEPR of the proposed construction is bounded above by $2$ for $2^k$ sub-carriers.  
\end{remark}
\begin{remark}
Although, we have taken CCCs from \cite{SaLiMa}, but authors in \cite{SaLiMa} didn't bound the column sequence PMEPR in proposed construction. In the proposed construction, we bound the maximum column sequence PMEPR of CCCs having flock size $p^k$ by $p$. 
\end{remark}
\section{More CC-ZCZ Code Sets}
In this section, another construction of CC-ZCZ code set is provided which are different from those generated from \emph{Theorem \ref{th1}}. Similar to CC-ZCZ constructed in \emph{Theorem \ref{th1}}, we also provide number of cosets of GRM codes,
number of CC-ZCZ code set in each coset, and Hamming distance but omit the proof as they are same as proof given in section IV. 

\par
 \begin{theorem}
    For a fixed $\mathbf{u_k}$, $\mathbf{C_{u_k}}$ is GCS as defined in \emph{Lemma \ref{l1}}. Let $\mathbf{C''_{u_k}}$ be $\mathbf{C_{u_k}}$ together with the permutation $\pi_{\beta}(n_\beta)=m-k+\beta$, for $\beta=1,2,\hdots,k$. Then $\mathbf{C''_{u_k}}$ is $(p^k,p^m,(p-1)p^{\pi_k(n_k-1)-1})$-Golay-ZCZ sequence set and hence $\mathbf{C''}=\{\mathbf{C''_{u_k}}:0\leq \mathbf{u_k}\leq p^k-1\}$ is $(p^k,p^k,p^m,(p-1)p^{\pi_k(n_k-1)-1})$-CC-ZCZ code set.\label{th2}  
  \end{theorem}
  
 \begin{corollary}
The number of coset representative $\mathcal{Q}_2$ in the second-order coset of the form $\mathcal{Q}_2+GRM_p(m,1)$ is given by 
\begin{equation}
    \sum_{n_1+n_2+\cdots+n_k=m}{{{m-k-1}\choose{n_1-1,n_2-1,\hdots,n_k-2}}(n_k-2)!\prod_{\beta=1}^{k-1}{(n_{\beta}-1)!}},
\end{equation}
where
\begin{equation}
        \mathcal{Q}_2=\sum_{\beta=1}^{k}{\sum_{\gamma=1}^{n_{\beta}-1}{x_{\pi_{\beta}(\gamma)}x_{\pi_{\beta}(\gamma+1)}}} \label{eq:80}.
\end{equation}
\end{corollary}

\begin{corollary}
Let $\mathcal{Q}_2+GRM_p(m,1)$ be any second-order coset
with coset representative $\mathcal{Q}_2$ as defined in \eqref{eq:80} and let $\pi_{k}(n_k-1) =m-k$. Then coset $\mathcal{Q}_2+GRM_p(m,1)$ consists of $p^{m-k+1}$ and $p^{m+1}$ distinct $p$-ary $(p^k,p^k,p^m,(p-1)p^{m-k-1})$-CC-ZCZ code set and $(p^k,p^m,(p-1)p^{m-k-1})$-Golay-ZCZ sequence set respectively.
\end{corollary}

\begin{corollary}
  The minimum Hamming distance of $(p^k,p^m,(p-1)p^{m-k-1})$-Golay-ZCZ sequence sets which are constructed in \emph{Theorem \ref{th2}} is $(p-1)p^{m-1}$.
\end{corollary}
{\color{black}
\section{Comparison with Existing Works}
Since the construction of CC-ZCZ is not available in the literature therefore in this section, 
the contributed Golay-ZCZ sequences, ZCZ sequences, and PMEPR of Golay-ZCZ sequences and CCCs are compared with existing works along with comparison tables.
\subsection{Comparison with existing constructions of Golay-ZCZ Sequence Set}
\subsubsection{Comparison with \cite{WaHu,ChWu}}
In \cite{ChWu}, authors proposed the construction of $(2^k,2^m,2^{\pi_1(2)-1})$-Golay-ZCZ sequence sets. In our construction (\emph{Theorem} \ref{th1}), if we take $p=2$ then the parameters of Golay-ZCZ sequence sets in \cite{ChWu} appear as special case of our parameters of Golay-ZCZ sequence sets. Similarly, in \cite{WaHu}, parameters of proposed $(2^k,2^m,2^{\pi_k(n_k)-1})$-Golay-ZCZ sequence sets appear as special case of ours for $p=2$ in \emph{Theorem} \ref{th2}.
\subsubsection{Comparison with \cite{Gu2021asymptotically,GuZh}}
In \cite{Gu2021asymptotically}, authors proposed an indirect construction of $(M,(M-1)N,M^2N)$-Golay-ZCZ sequence set with the help of $(M,M,N)$-CCC. This construction requires CCC as a seed code to construct Golay-ZCZ sequence set which makes it indirect. Furthermore, in the proposed construction, we are getting Golay-ZCZ sequence set through a multivariable function which makes our construction direct and advantageous over \cite{Gu2021asymptotically}. In \cite{GuZh}, authors proposed an indirect construction of Golay-ZCZ complementary pair of length $4N$ using GCPs of length $N$, having ZCZ width $N + 1$. This construction can produce only GCPs but not GCSs. Moreover, length of sequence is also dependent on length of GCP because the construction is indirect. On the other hand, the proposed construction is direct which is based on multivariable functions. Furthermore, the proposed construction can produce GCPs (for $p=2$) as well as GCSs (for $p\neq 2$).    \\
\begin{center}
\begin{table}{}
\resizebox{\textwidth}{!}{
\begin{threeparttable}
\caption{Comparision with \cite{WaHu,ChWu,Gu2021asymptotically,GuZh}}
\tiny
    \begin{tabular}{|c|c|m{5cm}|c|m{2cm}|} \hline
       \textbf{Author, Reference}  & \textbf{Based on} & \textbf{Parameter} (K,L,Z) \textbf{\& Constraints} & \textbf{Direct/Indirect} &
       \textbf{Optimality}\\ \hline
        Chen \emph{et. al}, \cite{ChWu}& GBF of order $2$ & $(2^k,2^m,2^{\pi_1{(2)}-1}),~\pi_\alpha(1)=m-\alpha+1$, $1\leq\beta\leq k$ & Direct & Optimal for binary case\\\hline
       Wang \emph{et. al}, \cite{WaHu}& GBF of order $2$& $(2^k,2^m,2^{\pi_k{(n_k-1)}-1}),~\pi_\alpha(n_\alpha)=m-k+\alpha$, $1\leq\beta\leq k$ & Direct & Optimal for binary case\\\hline
       
       Gu \emph{et. al}, \cite{Gu2021asymptotically}& $(M,M,N)$-CCC & $(M,(M-1)N,M^2N)$ & Indirect & Optimal for binary case\\\hline
       
       Gu \emph{et. al}, \cite{GuZh}& Concatenation of GCP & $(2,4N,N), ~ N$ is a positive integer & Indirect & Asymptotically optimal\\\hline
        
        Theorem 1 & Second-order multivariable functions & $(p^k,p^m,(p-1)p^{\pi_1(2)-1}),~p$ is prime and $\pi_\beta(1)=m-\beta+1,$ $1\leq\beta\leq k$ & Direct & Optimal for binary case else asymptotically optimal\\\hline
         Theorem 2 & Second-order multivariable functions & $(p^k,p^m,(p-1)p^{\pi_k(n_k-1)-1}),~p$ is prime and $\pi_\beta(n_\beta)=m-k+\beta,$ $1\leq\beta\leq k$ & Direct & Optimal for binary case else asymptotically optimal\\\hline
    \end{tabular}\label{t2}
   \end{threeparttable}}
\end{table}
\end{center}
\vspace{-2.5cm}

A detailed comparison of contributed Golay-ZCZ sequences with existing works is provided in Table \ref{t2}.
\subsection{Comparison with existing constructions of ZCZ Sequences}
{\color{black}The constructions of ZCZ sequences based on RM codes or Boolean functions \cite{TaYu,LiGuPa,ChWu}, multivariable functions \cite{NiSu}, bent functions\cite{ZhPa}, and perfect non-linear function\cite{ZhDz} are direct constructions of ZCZ sequences available in the literature while many constructions need kernel sequences such as perfect sequences \cite{LiCh,Th,ToNaSu,ZhTa}. Although, constructions presented in \cite{TaYu,LiGuPa,ChWu} are direct but length of ZCZ sequences is limited to power-of-two only. In \cite{ZhPa} and\cite{ZhDz}, the length of ZCZ sequences is limited to  $N^2$ and $p^2$ respectively, where $p$ is an odd prime, and $N$ is a positive integer. Hence flexibility in length parameter is not adequate here. Furthermore, in \cite{NiSu}, authors provided a direct construction of ZCZ sequences having flexible length, i.e., $p^n$, where $n$ is a positive integer $\geq 3$ but it is unable to produce ZCZ sequences of length $p^2$. The proposed construction can generate ZCZ sequences of length $p^n$, where $n$ is a positive integer $\geq 2$. A detailed comparison is provided in  Table \ref{t3}}.

\begin{table}[]
\begin{threeparttable}
\caption{Comparison with \cite{TaFa},\cite{ChWu},\cite{Th}, \cite{LiGuPa}, \cite{ZhDz}, \cite{ZhPa}, and \cite{TaYu}\tnote{*}}
    \tiny
    \centering
    \begin{tabular}{|c|c|m{4cm}|c|m{4cm}|}
    \hline
       \textbf{Author, Reference}  & \textbf{Based on} & \textbf{Parameter ({$K$},$L,Z$)} & \textbf{Direct/Indirect} &
       \textbf{Optimality}\\ \hline
        Tang \emph{et. al}, \cite{TaFa} & MOGCS & $(2^n,2^{n+1}Z_{cz},Z_{cz}+1),\ n \geq 1$ & Indirect & Optimal\\\hline
        {Chen \emph{et. al}, \cite{ChWu}} &  {Generalised Boolean function} &  {$(2^k,2^{m},2^{m-k-1}),$ $m\geq 2,$ and $k\leq m-1$.} &  Direct &  {Optimal for binary case}\\ \hline
        Hayashi \emph{et. al}, \cite{Th} & Perfect sequences & $(2(2n+1),4(2k+1)(2n+1),4k+1),\ n\geq 1,\ k\geq 1$ & Indirect & Neither optimal nor almost optimal\\\hline
        Liu \emph{et. al}, \cite{LiGuPa} & Second-order Reed-Muller codes & $(2^{k+1},2^{n+k+2},2^n),k\geq 0,n\geq 1$ & Direct & {Optimal for binary case}\\ \hline
        Zhou \emph{et. al}, \cite{ZhDz} & Perfect non-linear function & $(p,p^2,p)$, $p$ is an odd prime &  {Direct} & Optimal\\ \hline
        Zhang \emph{et. al}, \cite{ZhPa} & Generalised Bent function & $(N,N^2,N)$, $N$ is positive integer. &  {Direct} & Optimal.\\
        \hline
        Tang \emph{et. al}, \cite{TaYu} & Generalised Boolean function & $(2^k,2^m,2^{m-k}-2^{m-k-z})$ $,k\geq 0,n\geq 1$ & Direct & optimal if $m=k+z$ and almost optimal if $m=k+z+1$\\ \hline
        
        Kumar \emph{et. al}, \cite{NiSu} & Multivariable functions & ${(p^t,p^{n+t+1},(p-1)p^n)},\ n,t\geq 1, p$ is prime {and $t\leq n$} & Direct & Optimal for binary case else asymptotically optimal\\\hline
        
        Theorem 1 and Theorem 2 & Multivariable functions & $(p^k,p^m,(p-1)p^{m-k-1}),~p$ is prime & Direct & Optimal for binary case else asymptotically optimal\\\hline
    \end{tabular}\label{t3}
    {
    \begin{tablenotes}
      \item[*] The parameters presented in table are according to the definition of ZCZ in the corresponding paper. However, changing the definition of ZCZ also changes the bound and hence it doesn't affect the optimality of ZCZ sequence sets.
    \end{tablenotes}
    }
    \end{threeparttable}
\end{table}

}

\subsection{Comparison with PMEPR of Existing Works.}
In this section, the proposed PMEPR of CC-ZCZ code set is compared with the PMEPR of existing constructions of CCCs, detailed comparison is provided in Table \ref{t4}. It can be observed from the Table \ref{t4} that if we put $p=2$ in our construction then, the parameter of CCCs presented in \cite{ArAk,chen2008complete,liu2014new,wu2018optimal} and Golay-ZCZ sequence sets presented in \cite{ChWu,WaHu} can be obtained from our construction. Furthermore, since PMEPR of our CC-ZCZ code set is bounded by 2 hence it has advantage over \cite{ArAk,chen2008complete,wu2018optimal,ChWu,WaHu} and PMEPR of our CC-ZCZ code set coincide with PMEPR of CCC in \cite{liu2014new}. Additionally, the authors in \cite{SaLiMa} presented a construction of CCCs but did not bound the column sequence PMEPR. In Section V, we tackled this problem and provided a bound for column sequence PMEPR of CCC. 
\begin{table}[]
\centering
\caption{Comparison of the PMEPR of construction with \cite{ArAk,chen2008complete,WaHu,ChWu,Gu2021asymptotically,GuZh,liu2014new,wu2018optimal}}
\resizebox{\textwidth}{!}{
\begin{threeparttable}
\begin{tabular}{|l|l|l|l|l|l|}
 \hline

     \textbf{Author, Ref.} &  \textbf{Parameters} & \textbf{Based on} & \textbf{Constraints}  &\makecell{\textbf{Column sequence}\\ \textbf{PMEPR is upper}\\ \textbf{bounded by} } 
     \\ \hline
    
        Rathinakumar \emph{et. al}, \cite{ArAk}  & $(2^{k+1},2^{k+1},2^m)$-CCC  & GBF of order $2$       &$m,k\in \mathbb{Z}^+,~m>1 $ & $2^{k+1}$  \\ \hline
         
        Chen \emph{et. al},  \cite{chen2008complete} & $(2^k,2^k,2^m)$-CCC &  GBF of order $>2 $  & $k,m\in \mathbb{Z}^+$, $m\geq1,k\leq m$     & $2^k$ \\ \hline
        
         Wang \emph{et. al} \cite{WaHu} & $(2^k,2^m,2^{m-k-1})$-Golay-ZCZ & GBF & $m,k\in \mathbb{Z}^+,1\leq k \leq m-1$ & $2^k$\\\hline
     
      Chen \emph{et. al} \cite{ChWu} & $(2^k,2^m,2^{m-k-1})$-Golay-ZCZ & GBF &  $m,k\in \mathbb{Z}^+,1\leq k\leq m-1$ & $2^k$\\\hline

       Gu \emph{et. al}, \cite{Gu2021asymptotically}& $(M,(M-1)N,M^2N)$-Golay-ZCZ& CCC & There exist a $(M,M,N)$-CCC & $M$ \\\hline
       
       Gu \emph{et. al} \cite{GuZh}& $(2,4N,N)$-Golay-ZCZ& Concatenation of GCP & $N$ is a positive integer &  $2$\\\hline

       Z. Liu, \cite{liu2014new}& $(2^{k+1},2^{k+1},2^m)$-CCC & GBF of order $>2$  & $m,k\in \mathbb{Z}^+,~m>1 $ & $2$  
     \\ \hline
     
     Wu \emph{et. al}, \cite{wu2018optimal} & $(2^k,2^k,2^m)$-CCC &  GBF of order $2$ & $m,k\in \mathbb{Z}^+$, $m\geq3,1\leq k \leq m$ & $2^k$\\ \hline 
     

 This paper\tnote{*}  & $(p^k,p^k,(p-1)p^{m-k-1},p^m)$-CC-ZCZ &  MVF of order $2$     & $k,m\in \mathbb{Z}^+$, $p$ is prime  & $p$      \\ \hline 
 
\end{tabular}

 \begin{tablenotes}
   \item[*] In this paper CCCs are directly taken from \cite{SaLiMa} but we have bounded the column sequence PMEPR of these CCCs.
 \end{tablenotes}
 \end{threeparttable}}\label{t4}
\end{table}


\section{Conclusion}
In this paper, we {\color{black}settled open problem provided by Gong. \emph{et al.}} by providing a direct  construction of $(p^k,p^k,p^m,(p-1)p^{\pi_1(2)-1})$-CC-ZCZ code set and $(p^k,p^m,(p-1)p^{\pi_1(2)-1})$-Golay-ZCZ sequence set using multivariable functions. The proposed construction provides the non-power-of-two sequence length and larger ZCZ width. We are interested in larger ZCZ width therefore by substituting $\pi_1(2)=m-k$, $(p^k,p^m,(p-1)p^{m-k-1})$-CC-ZCZ code set is achieved, in which each Golay-ZCZ sequence set is asymptotically optimal for non-binary case else optimal, with respect to \emph{Tang-Fan-Matsufuji} Bound. We also established a relation between proposed CC-ZCZ code set and GRM codes and provided minimum Hamming distance of such codes. Moreover, the proposed construction provides tighter upper bound for column sequence PMEPR of proposed CC-ZCZ code set. The comparison of the proposed construction in the context of Golay-ZCZ sequences, ZCZ sequences, and PMEPR is provided. Further, relying on the existing literature, we propose two open problems as
\emph{
\begin{enumerate}
    \item Construction of CC-ZCZ code set and Golay-ZCZ sequence set having length in the form of product of different primes.
    \item Since each code in CC-ZCZ code set is ZCZ sequence set so we can treat CC-ZCZ code set as multiple ZCZ sequence set. Hence, it would be interesting to find inter-set ZCZ which is beneficial in multiuser environment to resist inter-cell interference caused by users from different cells in CDMA systems.
\end{enumerate}}
\begin{appendices}
\section{Proof of \textnormal{\textit{Theorem \ref{th1}}}}
We first state and prove a lemma which will later be used to prove the \emph{Theorem} \ref{th1}.
\begin{lemma}\label{l2}
   For two non-negative integers $i$ and $j$, let $(i_1,i_2,\hdots,i_m)$ and $(j_1,j_2,\hdots,j_m)$ be $p$-ary representation of $i$ and $j$ respectively. Let $f:\mathbb{Z}_p^m\rightarrow\mathbb{Z}_\lambda$ be a function as defined in \eqref{eq:12}. Now, we can write $a_{\mathbf{u_k}}^{\mathbf{v_k}}$ by using \eqref{eq:13} and \eqref{eq:14} as
   \begin{equation}
       a_{\mathbf{u_k}}^{\mathbf{v_k}}=f+\frac{\lambda}{p}\sum_{\beta=1}^{k}{x_{\pi_{\beta}{(1)}}u_{\beta}}+\frac{\lambda}{p}\sum_{\beta=1}^{k}{x_{\pi_{\beta}{(n_\beta)}}v_{\beta}}.\label{eq:16}
   \end{equation}
   Further, suppose we have $i_{\pi_\beta(1)}=j_{\pi_\beta(1)}~\forall~\beta=1,2,\hdots,k$. For a certain $\beta'\leq k$, let us assume $t$ be the smallest integer such that $i_{\pi_{\beta'}(t)}\neq j_{\pi_{\beta'}(t)}$. Let us define $i^\eta$ to be an integer whose vector representation with base $p$ is
   \begin{equation}
       (i_1,i_2,\hdots,i_{\pi_{\beta'}(t-1)}-\eta,\hdots,i_m),\label{eq:72}
   \end{equation}
   which differs from that of $i$ only at the position $\pi_{\beta'}(t-1)$
   and $\eta=1,2,\hdots,(p-1)$. Similarly, we define $j^\eta$ such that its vector representation with base $p$ is
   \begin{equation}
       (j_1,j_2,\hdots,j_{\pi_{\beta'}(t-1)}-\eta,\hdots,j_m).\label{eq:73}
   \end{equation}
  Then we have 
   \begin{equation}
       \sum_{\eta=1}^{p-1}{\omega_\lambda^{(a_{\mathbf{u_k}}^{\mathbf{v_k}})_{j^\eta}-(a_{\mathbf{u_k}}^{\mathbf{v_k}})_{i^\eta}}}+\omega_\lambda^{(a_{\mathbf{u_k}}^{\mathbf{v_k}})_{j}-(a_{\mathbf{u_k}}^{\mathbf{v_k}})_{i}}=0.\label{eq:18}
   \end{equation}
\end{lemma}
\begin{IEEEproof}
For notational convenience, we write $a_{\mathbf{u_k}}^{\mathbf{v_k}}=\mathbf{c}$. Now using \eqref{eq:12} and \eqref{eq:16}, we can write 
\begin{align}
    \mathbf{c}_{i^\eta}-\mathbf{c}_{i}=&f_{i^\eta}-f_{i}  \nonumber\\
    =&\frac{\lambda}{p}[i_{\pi_{\beta'}(t-2)}i_{\pi_{\beta'}(t-1)}^\eta+i_{\pi_{\beta'}(t-1)}^\eta i_{\pi_{\beta'}(t)}-i_{\pi_{\beta'}(t-2)}i_{\pi_{\beta'}(t-1)}^\eta-i_{\pi_{\beta'}(t-1)}i_{\pi_{\beta'}(t)}] \nonumber \\ 
    &+g_{\pi_{\beta'}(t-1)}[i_{\pi_{\beta'}(t-1)}^\eta-i_{\pi_{\beta'}(t-1)}] \nonumber\\
    =&[-\eta\frac{\lambda}{p} i_{\pi_{\beta'}(t)}-\eta\frac{\lambda}{p} i_{\pi_{\beta'}(t-2)}-\eta g_{\pi_{\beta'}(t-1)}] \nonumber \\
    =&-\eta[\frac{\lambda}{p}i_{\pi_{\beta'}(t)}+\frac{\lambda}{p}i_{\pi_{\beta'}(t-2)}+g_{\pi_{\beta'}(t-1)}].\label{eq:20}
\end{align}
Similarly,
\begin{equation}
    \mathbf{c}_{j^\eta}-\mathbf{c}_{j}=f_{j^\eta}-f_{j}=-\eta[\frac{\lambda}{p}j_{\pi_{\beta'}(t)}+\frac{\lambda}{p}j_{\pi_{\beta'}(t-2)}+g_{\pi_{\beta'}(t-1)}].\label{eq:21}
\end{equation}
From \eqref{eq:20} and \eqref{eq:21}, we have
\begin{equation}
   (\mathbf{c}_{i^\eta}-\mathbf{c}_{j^\eta})-(\mathbf{c}_{i}-\mathbf{c}_{j})=-\eta\frac{\lambda}{p}(i_{\pi_{\beta'}(t)}-j_{\pi_{\beta'}(t)}). \label{eq:22}
\end{equation}
Now, taking sum of $\omega_\lambda^{(\mathbf{c}_{i^\eta}-\mathbf{c}_{j^\eta})-(\mathbf{c}_{i}-\mathbf{c}_{j})}$ over $\eta$, we have
\begin{equation}
    \sum_{\eta=1}^{p-1}{\omega_\lambda^{\mathbf{c}_{i^\eta}-\mathbf{c}_{j^\eta}-(\mathbf{c}_{i}-\mathbf{c}_{j})}=\sum_{\eta=1}^{p-1}{\omega_\lambda^{\eta\frac{\lambda}{p}(j_{\pi_{\beta'}(t)-i_{\pi_{\beta'}(t)}})}}}=\sum_{\eta=1}^{p-1}{\omega_p^{\eta(j_{\pi_{\beta'}(t)-i_{\pi_{\beta'}(t)}})}}. \label{eq:23}
\end{equation}
It is given that $i_{\pi_{\beta'}(t)}\neq j_{\pi_{\beta'}(t)}$, implies that $i_{\pi_{\beta'}(t)}-j_{\pi_{\beta'}(t)}\neq0$ and hence RHS of \eqref{eq:23} is the sum of roots of polynomial $z^p-1=0$ except the root $z=1$. Hence
\begin{align}
    \sum_{\eta=1}^{p-1}{\omega_p^{\eta(j_{\pi_{\beta'}(t)-i_{\pi_{\beta'}(t)}})}}=-1. \label{eq:24}
\end{align}
Therefore, from \eqref{eq:23} and \eqref{eq:24}, we have
\begin{equation}
    \sum_{\eta=1}^{p-1}{\omega_\lambda^{\mathbf{c}_{i^\eta}-\mathbf{c}_{j^\eta}-(\mathbf{c}_{i}-\mathbf{c}_{j})}}=-1,\label{eq:25}
\end{equation}
which further implies that
\begin{equation}
    \sum_{\eta=1}^{p-1}{\omega_\lambda^{\mathbf{c}_{i^\eta}-\mathbf{c}_{j^\eta}}+\omega_\lambda^{(\mathbf{c}_{i}-\mathbf{c}_{j})}}=0.
\end{equation}
\end{IEEEproof}
  \begin{IEEEproof}[Proof of Theorem \ref{th1}]
      We need to show that for a fixed $\mathbf{u_k}$, $\mathbf{C_{u_k}}$ is a $(p^k,(p-1)p^{\pi_1(2)-1},p^m)$-ZCZ. Except the ZCZ width all the parameters are directly inherited from \emph{Lemma \ref{l1}}. So, we only need to show that the ZCZ width is $(p-1)p^{\pi_1(2)-1}$. Let $\mathbf{c}\in\mathbf{C_{u_k}}$, by \eqref{eq:15}, $\mathbf{c}=\psi(a_{\mathbf{u_k}}^{\mathbf{v_{k_1}}})$ for some $0\leq\mathbf{v_{k_1}}\leq p^k-1$. First, we find the PACF of $\mathbf{c}$ and show that for $0<\tau\leq(p-1)p^{\pi_1(2)}-1$,
      \begin{equation}
         \mathcal{P}(\mathbf{c})(\tau)=\sum_{i=0}^{L-1}{\omega_\lambda^{\mathbf{c}_{(i+\tau)mod~L}-\mathbf{c}_i}}=0,\label{eq:27}
      \end{equation}
      where $L$ is the length of the sequence, i.e., $L=p^m$.
     For any integer $i$, let us denote another integer $j=(i+\tau)mod~L$. Then we consider two cases and demonstrate that for each pair $(i,j)$ there exist other $(p-1)$ pairs $(i^\eta,j^\eta),~\eta=1,2,\hdots,p-1$ such that
     \begin{equation}
         \sum_{\eta=1}^{p-1}{\omega_\lambda^{\mathbf{c}_{i^\eta}-\mathbf{c}_{j^\eta}}+\omega_\lambda^{(\mathbf{c}_{i}-\mathbf{c}_{j})}}=0,\label{eq:28}
     \end{equation}
    in each case.
    \begin{case}[$i_{\pi_{1}(2)}=j_{\pi_{1}(2)}$]
      In this case, we have $i_{\pi_{\beta}(1)}=j_{\pi_{\beta}(1)},~\forall \beta=1,2,\hdots,k$. For the contrary, suppose this is not true. Then, assume that $\bar{\beta}$ is the largest integer such that $i_{\pi_{\bar{\beta}}(1)}\neq j_{\pi_{\bar{\beta}}(1)}$. For ease of presentation, let $d=\pi_{\bar{\beta}}(1)$, now if $j_d>i_d$, we have
      \begin{align}
          \tau=j-i=&\sum_{s=1}^{d}{(j_s-i_s)p^{s-1}} \nonumber\\
          =&(j_d-i_d)p^{d-1}+\sum_{s=1,s\neq \pi_{1}(2)}^{d-1}{(j_s-i_s)p^{s-1}}\nonumber\\
          \geq& (j_d-i_d)p^{d-1}-(p-1)\sum_{s=1}^{d-1}{p^{s-1}}+(p-1)p^{\pi_{1}(2)-1}\nonumber
          \end{align}
          \begin{align}
          =& (j_d-i_d)p^{d-1}-(p-1)\Big[\frac{p^{d-1}-1}{p-1}\Big]+(p-1)p^{\pi_{1}(2)-1}\nonumber\\
          =& (j_d-i_d-1)p^{d-1}+1+(p-1)p^{\pi_{1}(2)-1}>(p-1)p^{\pi_{1}(2)-1}.\label{eq:29}
      \end{align}
      Hence \eqref{eq:29} implies that $\tau>(p-1)p^{\pi_{1}(2)-1}$ which is a contradiction. Similarly, if $j_d<i_d$, then
      \begin{align}
      \tau&=j-i+p^m=\sum_{s=1}^{d}{(j_s-i_s)p^{s-1}}+p^m\nonumber\\
      &=(j_d-i_d)p^{m-\bar{\beta}}+p^m+\sum_{s=1,s\neq\pi_{1}(2)}^{d-1}{(j_s-i_s)p^{s-1}}\nonumber\\
      &=(j_d-i_d+p^{\bar{\beta}})p^{m-\bar{\beta}}+\sum_{s=1,s\neq\pi_{1}(2)}^{d-1}{(j_s-i_s)p^{s-1}}\nonumber\\
      &\geq (j_d-i_d+p^{\bar{\beta}})p^{m-\bar{\beta}}-(p-1)\sum_{s=1}^{d-1}{p^{s-1}}+(p-1)p^{\pi_{1}(2)-1}\nonumber
      \end{align}
      \begin{align}
      =& (j_d-i_d+p^{\bar{\beta}}-1)p^{m-\bar{\beta}}+1+(p-1)p^{\pi_{1}(2)-1}>(p-1)p^{\pi_{1}(2)-1}.
      \end{align}
      Again we got a contradiction. Hence $i_{\pi_{\beta}(1)}=j_{\pi_{\beta}(1)}~\forall \beta=1,2,\hdots,k$. Now without loss of generality, we assume that there exist a positive integer $\beta'\leq k$ such that $i_{\pi_{\beta}(r)}=j_{\pi_{\beta}(r)},~\forall \beta=1,2,\hdots,\beta'-1$ and $r=1,2,\hdots,n_{\beta}$. Assume $t$ be the smallest integer with $i_{\pi_{\beta'}(t)}\neq j_{\pi_{\beta'}(t)}$. Now, let us define $i^\eta$ and $j^\eta$ same as in \eqref{eq:72} and \eqref{eq:73} respectively. Then it can easily be obtained that $j^\eta=(i^\eta+\tau)mod~L$ and hence using \emph{Lemma} \ref{l2}, we get \eqref{eq:28}.
    \end{case}
    \begin{case}[$i_{\pi_{1}(2)}\neq j_{\pi_{1}(2)}$]
      In this case, let $i^\eta$ and $j^\eta$ are modified from $i$ and $j$ by changing only last bit of $i$ and $j$ as $i_{m}^\eta=i_{m}-\eta$ and $j_{m}^\eta=j_{m}-\eta$. Then
      \begin{align}
        \mathbf{c}_{i^\eta}-\mathbf{c}_{i}&=f_{i^\eta}-f_{j}+\frac{\lambda}{p}(i_m-\eta-i_m)u_1\nonumber\\
        &=\frac{\lambda}{p}\left[i_m^\eta i_{\pi_1(2)}-i_mi_{\pi_1(2)}\right]+g_{m}(i_m-\eta-i_m)+\frac{\lambda}{p}(i_m-\eta-i_m)u_1\nonumber\\
        &=-\eta\left[\frac{\lambda}{p}i_{\pi_1(2)}+g_m+\frac{\lambda}{p}u_1\right].\label{eq:31}
      \end{align}
      Similarly,
      \begin{equation}
          \mathbf{c}_{j^\eta}-\mathbf{c}_{j}=-\eta\left[\frac{\lambda}{p}j_{\pi_1(2)}+g_m+\frac{\lambda}{p}u_1\right].\label{eq:32}
      \end{equation}
      By subtracting \eqref{eq:32} from \eqref{eq:31}, we get
      \begin{equation}
          (\mathbf{c}_{i^\eta}-\mathbf{c}_{j^\eta})-(\mathbf{c}_{i}-\mathbf{c}_{j})=-\eta\frac{\lambda}{p}(i_{\pi_{1}(2)}-j_{\pi_{1}(2)}). \label{eq:33}
      \end{equation}
      Now following the same steps as in \eqref{eq:22}, \eqref{eq:23}, \eqref{eq:24}, and \eqref{eq:25}, we get
      \begin{equation}
         \sum_{\eta=1}^{p-1}{\omega_\lambda^{\mathbf{c}_{i^\eta}-\mathbf{c}_{j^\eta}}+\omega_\lambda^{(\mathbf{c}_{i}-\mathbf{c}_{j})}}=0.
     \end{equation}
      Till now we have proved that for $0<\tau\leq(p-1)p^{\pi_1(2)-1}$, the value of $\mathcal{P}(\mathbf{c})(\tau)=0$. Now in the rest of the proof, we will prove that for any $0\leq\tau\leq(p-1)p^{\pi_1(2)-1}$, the PCCF of any two different sequences in $\mathbf{C_{u_k}}$ is zero. For that let $0\leq\boldsymbol{\gamma_k},\boldsymbol{\delta_k}\leq p^k-1$ such that $\psi(a_{\mathbf{u_k}}^{\boldsymbol{\gamma_k}}),\psi(a_{\mathbf{u_k}}^{\boldsymbol{\delta_k}})\in\mathbf{C_{u_k}}$. Again for notational convenience, we denote $\psi(a_{\mathbf{u_k}}^{\boldsymbol{\gamma_k}})$ and $\psi(a_{\mathbf{u_k}}^{\boldsymbol{\delta_k}})$ by $\mathbf{b}$ and $\mathbf{c}$ respectively. Then, we need to prove that for $0\leq\tau\leq(p-1)p^{\pi_1(2)-1}$,
      \begin{equation}
         \mathcal{P}(\mathbf{c},\mathbf{b})(\tau)=\sum_{i=0}^{L-1}{\omega_\lambda^{\mathbf{c}_{(i+\tau)mod~L}-\mathbf{b}_i}}=0.\label{eq:34}
      \end{equation}
Let $j=(i+\tau)mod~L$. Following similar arguments as in \emph{Case} $1$ and \emph{Case} $2$, for any pair $(i,j)$, we can find other pairs $(i^\eta,j^\eta),~\eta=1,2,\hdots,k-1$
such that
\begin{equation}
    \sum_{\eta=1}^{p-1}{\omega_\lambda^{\mathbf{c}_{i^\eta}-\mathbf{b}_{j^\eta}}+\omega_\lambda^{(\mathbf{c}_{i}-\mathbf{b}_{j})}}=0,
\end{equation}
for $\tau\neq0$, where $j^\eta=(i^\eta+\tau)mod~L$. Therefore, we can obtain that \eqref{eq:34} holds for $\tau\neq0$. Now, it remains to prove that,
\begin{equation}
         \mathcal{P}(\mathbf{c},\mathbf{b})(0)=\sum_{i=0}^{L-1}{\omega_\lambda^{\mathbf{c}_{i}-\mathbf{b}_i}}=0.\label{eq:35}
\end{equation}
Taking \eqref{eq:16} into consideration, let $\mathbf{c}-\mathbf{b}=\frac{\lambda}{p}.\mathbf{d}$. Then $\mathbf{d}$ is a non-zero codeword in $GRM_p(1,m)$. Now let $\mathbf{d}=(d_1,d_2,\hdots,d_{p^m})$ and hence $\mathbf{d}$ can be written as linear combination of $\theta(x_1),\theta(x_2),$ $\hdots ,\theta(x_m)$ as
$\mathbf{d}=c_1\cdot\theta(x_1)+c_2\cdot\theta(x_2)+\cdots+c_m\cdot\theta(x_m)$, where $c_i\in \mathbb{Z}_p,~1\leq i\leq m$. For each $i$, $\theta(x_i)$ contains each element of the set $\{0,1,\hdots,p-1\}$, $p^{m-1}$ times. Hence $\mathbf{d}$ will also contains each element of the set $\{0,1,\hdots,p-1\}$, $p^{m-1}$ times. Now \eqref{eq:35} can be written as
 \begin{equation}
         \mathcal{P}(\mathbf{c},\mathbf{b})(0)=\sum_{i=0}^{L-1}{\omega_\lambda^{\frac{\lambda}{p}\mathbf{d}_{i}}}=\sum_{i=0}^{L-1}{\omega_\lambda^{\frac{\lambda}{p}\mathbf{d}_{i}}}=\sum_{i=0}^{L-1}{\omega_p^{\mathbf{d}_{i}}}=0.
\end{equation}
Hence the \emph{Theorem} \ref{th1} is proved.
    \end{case}
  \end{IEEEproof}
  \end{appendices}
\bibliographystyle{IEEEtran}
\bibliography{CCC_ZCZ}
\end{document}